\documentclass{emulateapj} \usepackage{apjfonts}

\newcommand{\Msun}{M_{\odot}}

\shorttitle{Where are the ``Missing'' Galactic Baryons?}
\shortauthors{Sommer-Larsen}

\begin{document}


\title{Where are the ``Missing'' Galactic Baryons?}


\author{Jesper Sommer-Larsen}
\affil{Dark Cosmology Centre, Niels Bohr Institute, Juliane Maries Vej 30, DK-2100 Copenhagen {\O}, Denmark}
\email{jslarsen@tac.dk}


\begin{abstract} 
Based on 19 high-resolution N-body/gasdynamical galaxy  formation 
simulations in the $\Lambda$CDM cosmology it is shown, that for a galaxy
like the Milky Way, in addition to the baryonic mass of the galaxy itself,
about 70\% extra baryonic mass should reside
{\it around} the galaxy (inside of the virial radius), chiefly in
the form of hot gas. Averaging over the entire field galaxy population, this
``external'' component amounts to 64-85\% of the baryonic mass of the
population itself. These results are supported by the recent detection of
very extended, soft X-ray emission from the halo of the quiescent, massive disk
galaxy NGC~5746. Some of the hot gas may, by
thermal instability, have condensed into mainly pressure supported, warm 
clouds, similar to the
Galactic High Velocity Clouds (HVCs). Based on an ultra-high resolution
cosmological test simulation of a Milky Way like galaxy (with a gas particle 
mass and gravity softening length of only 7.6x10$^3$~$h^{-1}\Msun$ and 
83$h^{-1}$pc,
respectively), it is argued, that the hot gas phase dominates over the warm
gas phase, in the halo.
Finally, an origin of HVCs as ``leftovers'' from filamentary, ``cold''
accretion events, mainly occurring early in the history of galaxies, 
is proposed. 
\end{abstract}




\section{Introduction}
In the $\Lambda$CDM cosmology a disk galaxy like the Milky Way will have
a virial mass of $\sim$8x10$^{11}~\Msun$. For a universal baryon
fraction $f_b\sim$~0.15, one would expect the baryonic mass of the
Milky Way to be $\sim$1.2x10$^{11}~\Msun$, assuming that the bulk
of the baryonic material inside of $r_{\rm{vir}}$ ($\sim$~250 kpc) is
deposited onto the central galaxy. However, the baryonic mass (stars +
cold gas) of the Milky Way is found to be just $\sim$6x10$^{10}~\Msun$
(e.g., Dehnen \& Binney 1998; Sommer-Larsen \& Dolgov 2001), 
so an amount of baryonic mass, as large as the actual mass of the 
Milky Way itself, appears to be ``missing'' (e.g., Silk 2004, 
Maller \& Bullock 2004).

On the other hand, on theoretical grounds it has been known for long
that galaxies (not only in groups and clusters, but also in the ``field'')
should be embedded in extended haloes of hot gas (e.g., White \& Frenk 1991; 
Sommer-Larsen 1991), and evidence for hot, dilute gas in the Galactic halo is
quite strong (e.g., Sembach et al.~2003). 
Searches for X-ray emission from the haloes of external, quiescent disk 
galaxies have
until very recently proved unsuccessful (as opposed to star-burst galaxies,
e.g., Strickland et al.~2002), which has been taken as an indication
that the baryonic mass of such hot haloes is insignificant, due to, e.g.,
strong AGN driven hot gas outflows at some point during galaxy formation
(e.g., Benson et al. 2000).

Based on recent cosmological simulations of disk galaxy formation (not
invoking violent AGN feedback) \cite{T.02} showed, however, that the
X-ray null detections were to be expected. Moreover, they showed that
the X-ray luminosity of disk galaxy haloes is expected to be a very steep 
function of the characteristic circular speed, roughly as $L_X \propto V_c^7$.
Very recently \cite{P.06} observed the massive, quiescent, isolated and 
edge-on, disk 
galaxy NGC5746 ($V_c=305\pm7$~km/s), and detected hot halo soft 
($kT\sim$0.4 keV) X-ray emission at the level predicted by the numerical
models (see also Rasmussen et al. 2006 for more detail). 
As the mass of hot gas inside of $r_{\rm{vir}}$ for such a galaxy
is predicted to be $\sim$~80\% of that of the central galaxy, and the total
``external'' baryonic mass fraction $\sim$~110\% (section 3), it is
clearly of interest to estimate the global external baryonic mass-fraction,
averaged over the entire field galaxy population. This is the aim of this
Letter: in sec. 2 the code and simulations are briefly described,
the results obtained are presented in sec.3, and discussed in sec.4.
Issues like formation of warm, mainly
pressure supported clouds (``High Velocity Clouds'') through thermal 
instability of hot halo gas, are also addressed\footnote{For images, 
and an HVC animation, see http://www.tac.dk/\~~\hspace{-1.4mm}jslarsen/HVC}.

\vspace*{-0.2cm}
\section{THE CODE AND SIMULATIONS}
The code used for the simulations is a significantly improved version of
the TreeSPH code, which has been used previously for galaxy formation 
simulations (Sommer-Larsen, G\"otz \& Portinari 2003, SLGP03).
The main improvements over the previous version are:
(1) The ``conservative'' entropy 
equation solving scheme suggested by \cite{SH02} has been adopted. 
(2) Non-instantaneous gas recycling and chemical evolution, tracing
10 elements (H, He, C, N, O, Mg, Si, S, Ca and Fe), has been incorporated
in the code following Lia et~al.\ (2002a,b); the algorithm includes 
supernov\ae\ of type II and type Ia, and mass loss from stars of all masses.
(3) Atomic radiative cooling depending both on the metal abundance
of the gas and on the meta--galactic UV field, modeled after Haardt
\& Madau (1996) is invoked, as well as simplified treatment
of radiative transfer, switching off the UV field where the gas
becomes optically thick to Lyman limit photons on scales of $\sim$ 1~kpc.

For the present project, the formation and evolution of 15 individual
galaxies, known from previous work to become disk galaxies at $z$=0,
was simulated with the above, significantly improved, TreeSPH code.
At least two different numerical resolutions were used to simulate
each galaxy. Moreover, many of the galaxies were also simulated with different
physical prescriptions for the early ($z\ga4$) star-bursts (and related
SNII driven energy feedback) found previously to be required, in order
to produce realistic disk galaxies. The 15 galaxies were selected 
from a cosmological LCDM simulation (see below) to
represent ``field'' galaxies, by requiring each galaxy to be at least
1 Mpc from any galaxy group, and at least 0.5 Mpc away from any larger
galaxy, at $z$=0. The disk galaxies span a range of characteristic 
circular speeds of $V_c \sim 100-330$ km/s, and a range of virial
masses of 6x10$^{10}$ to 3x10$^{12} \Msun$.

The galaxies (galaxy DM haloes) were selected from a
cosmological, DM-only simulation of box-length 10 $h^{-1}$Mpc
(comoving), and starting redshift $z_i$=39.  
The adopted cosmology was the flat $\Lambda$CDM model, with
($\Omega_M$, $\Omega_{\Lambda}$)=(0.3,0.7).

Mass and force resolution was increased in Lagrangian regions enclosing the 
galaxies, and in these regions all DM particles were split into a DM particle
and a gas (SPH) particle according to an adopted universal baryon fraction of
$f_b$=0.15, in line with recent estimates. In this paper, only results of
high-resolution simulations, consisting of at least 1.5x10$^5$ 
SPH+DM particles,
will be presented (with typical numbers in the range 2-3x10$^5$). 
Comparison
to simulations of lower and very high resolution, respectively, will be
discussed, though. For galaxies of $V_c < 150$ km/s, $m_{\rm{gas}}$=$m_*$=
9.1x10$^4$ and $m_{\rm{DM}}$=5.2x10$^5$ $h^{-1}$M$_{\odot}$.
Moreover, gravitational (spline) 
softening lengths of $\epsilon_{\rm{gas}}$=$\epsilon_*$=190 and 
$\epsilon_{\rm{DM}}$=340 $h^{-1}$pc, respectively, were adopted. 
For galaxies of $V_c \ge 150$ km/s, $m_{\rm{gas}}$=$m_*$=
7.3x10$^5$ and $m_{\rm{DM}}$=4.2x10$^6$ $h^{-1}$M$_{\odot}$, and
$\epsilon_{\rm{gas}}$=$\epsilon_*$=380 and 
$\epsilon_{\rm{DM}}$=680 $h^{-1}$pc. In addition, two galaxies of
$V_c$=180 and 244 km/s, respectively, were re-simulated with the
smaller particle masses and gravity softenings above. These very
high resolution runs consisted of 1.2 and 2.2 million particles. 
The
gravity softening lengths were fixed in physical coordinates from $z$=6
to $z$=0, and in comoving coordinates at earlier times.

A Kroupa IMF was used in the simulations, and early rapid and 
self-propagating star-formation (sometimes dubbed ``positive feedback'')
was invoked (SLGP03). Finally, in order to enable some reuse of previous
work, values of $h$=0.65 and $\sigma_8$=1.0 were employed in the 
cosmological simulations. These values are slightly different from the
$h \simeq 0.7$ and $\sigma_8 \simeq 0.9$, currently favored. To check
the effect of this, one simulation was undertaken with $h=0.7$ and 
$\sigma_8=0.9$.
\begin{figure}[t]
\begin{center}
\resizebox{7.8cm}{!}{\includegraphics{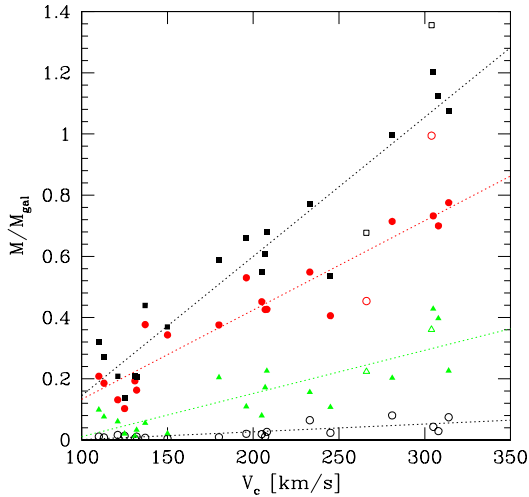}}\vspace*{-0.4cm}
\end{center}%
\vspace{-0.7cm}
\caption{For 19 high-resolution disk galaxy simulations are shown: total
external baryon mass-fractions ({\it solid squares}), and contributions
from hot gas ({\it solid circles}), satellite galaxies + outer halo stars 
({\it solid triangles}), and inner halo stars ({\it open circles}); linear
fits to the various components are shown by dotted lines. Also
shown, by open symbols, are results for two elliptical galaxies (inner halo 
stars included in the central galaxy, hence not shown).}
\end{figure} 

\section{RESULTS}
A detailed analysis of the simulation results, in general, will be given
elsewhere, e.g., \cite{PSL06};
here emphasis will be just on ``external'' baryonic mass-fractions:
For each disk galaxy simulation the distribution of baryonic mass, at $z$=0
and inside of the virial radius, was classified as follows: 1) disk+bulge stars
($R<20(V_c/220 \rm{km/s})$~kpc, $|z|<5(V_c/220 \rm{km/s})$kpc), 2) inner
halo stars ($r<20(V_c/220 \rm{km/s})$~kpc; not in the above region), 
3) stellar satellites/outer halo ($r<r_{\rm{vir}}$, and not in the above 
regions), 4) gas in the galaxy (mainly ``cold'': $T\le3$x$10^4$~K; 
same region as for the disk+bulge stars), 5) outer hot gas ($T>3$x$10^4$~K; 
$r<r_{\rm{vir}}$, outside disk+bulge region), and 6) outer cold gas 
($r<r_{\rm{vir}}$; outside disk+bulge region).   
The hot gas is typically at $T \sim T_{\rm{vir}}$, and hence, in general, 
much hotter than 3x10$^4$~K. 

The simple estimate of the galactic baryonic mass would be the combined
mass of components 1 and 4. In Fig.1 is shown, for 19 high-resolution
disk galaxy simulations, the mass of the other components relative to
this baryonic galaxy mass. Components 3 and 6 have been combined, since
most of the ``external'' cold gas is located in satellite galaxies. As
can be seen, the importance of the three resulting ``external'' 
components increases with $V_c$: at $V_c\sim$100 km/s, the external
baryonic mass-fraction is only about 20\%, at $V_c\sim$220 km/s (like
the Milky Way) this fraction increases to $\sim$70\%, and at 
$V_c\sim$300 km/s it is about a factor 1.1. At all $V_c$, the hot gas
is the dominant external component.
\begin{figure}[t]
\begin{center}
\resizebox{7.8cm}{!}{\includegraphics{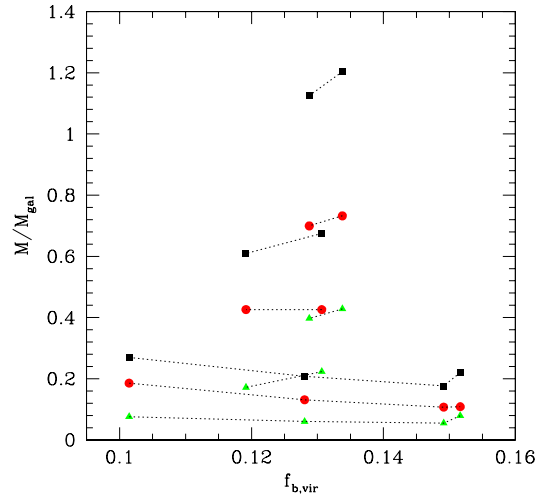}}\vspace*{-0.4cm}
\end{center}%
\vspace{-0.7cm}
\caption{External fractions for 3 galaxies, simulated at high resolution
with various star-formation and feedback prescriptions, versus total baryonic
mass fractions within the virial radius (symbols as in
Fig.1; results for the same galaxy are connected by lines). For a
given galaxy, as SNII feedback weakens, $f_{\rm{b,vir}}$ systematically 
increases.}
\end{figure}
Before inferring the implications of these results, their dependence 
on the physics invoked in the simulations, as well as on numerical
resolution, should be assessed. To this end is compared in Fig.2 the results
of various simulations of three galaxies, all run at high resolution, with
different prescriptions for SNII feedback. For a small galaxy, of 
$V_c\sim$110 km/s, is shown results for four different simulations: two
with early, self-propagating star-formation (SPSF), and associated strong 
feedback (for two different threshold densities $n_{\rm{H,e,low}}$=0.1 and 
0.25 cm$^{-3}$; SLGP03), 
one with early fast star-formation, but no SPSF (this reduces the early
feedback considerably), and one with no early fast star-formation at all
(this corresponds to the very weak feedback case). Fig.2 shows that a)
the stronger the feedback the smaller is the resulting baryon fraction
inside of the virial radius, and b) the quantities shown in Fig.1 are
quite robust to such a dramatic change of feedback physics (the two lower
feedback simulations result in galaxies with smaller specific angular momenta
and stellar disks, which are kinematically too hot compared to observed
disks --- hence the results for these have not been included in Fig.1). 
Also shown are data for SPSF simulations, with the above two 
thresholds, for $V_c\sim$205 km/s and a $V_c\sim$305 km/s galaxies,
respectively. As can be seen, these results are quite robust as well.

To test for resolution effects, results for 7 galaxies with $V_c$ in the
range 170-250 km/s, run at low resolution (the main one of SLGP03)
and high resolution, but identical feedback descriptions, were compared. 
It was found that the hot gas fraction decreased by 28$\pm$4\%, and the total
``external'' baryonic mass fraction by 18$\pm$6\%, increasing the mass
resolution by a factor of eight, and the force resolution by a factor
of two. For a single galaxy of $V_c\sim$305, the corresponding numbers
were 23 and 13\%. To test the effects of going to very high resolution,
results for two galaxies, of $V_c$=180 and 244 km/s, were
compared at $z\sim$~1 and 2, respectively (results at $z$=0 were not available,
but at both redshifts gas cool-out and star formation is very well underway). 
It was found that in going from
high to very high resolution, the hot gas fraction is reduced only by about
5\% more, and the total external baryon fraction did not change at all. 
On basis
of the above, only simulations run at high resolution were used in the 
present analysis.
\begin{figure}[t]
\begin{center}
\vspace*{-0.5cm}\resizebox{7.8cm}{!}{\includegraphics{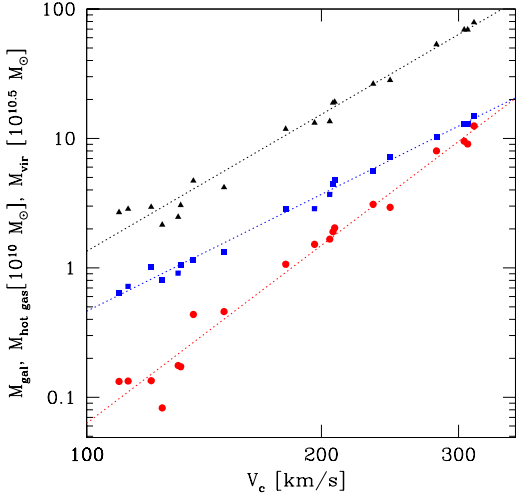}}\vspace*{-0.5cm}
\end{center}%
\vspace{-0.5cm}
\caption{Galaxy baryonic mass (disk+bulge+cold gas),  
({\it solid squares}), total virial mass ({\it solid triangles}), and
hot gas mass ({\it solid circles}) versus $V_c$. Dotted lines show power 
law fits.}
\end{figure}

\section{Discussion}
To estimate the global importance of the external baryons in the mass
budget of disk galaxies,
the global distribution of disk galaxy circular velocities (the ``velocity
function'') is required. Galaxy velocity functions have been estimated by, 
e.g., \cite{G.00} and \cite{D.04}. Most published velocity functions
refer to the general morphological mix of disk, lenticular and elliptical
galaxies, in the field or in clusters, but \cite{G.00} give an expression
for field disk galaxies, viz.
\begin{equation}
\Psi_d(V_c)~dV_c = \Psi_{d,*} \left(\frac{V_c}{V_{c,*}}\right)^{\beta}
\exp\left[-\left(\frac{V_c}{V_{c,*}}\right)^{n}\right] ~\frac{dV_c}{V_{c,*}} ~~,
\end{equation}
where $\Psi_{d,*}=2.0\pm0.4~10^{-2}~h^3$ Mpc$^{-3}$, $\beta=1.3\pm0.18$,
$n$=2.5 and $V_{c,*}=247\pm7$ km/s. Next, one has to determine the relation
between $V_c$ and the combined baryonic mass of the disk+bulge stars and 
gas in the galaxy. In Fig.3 this relation is shown for the 19 high
resolution simulations. The relation is very well fitted by a power law 
\begin{equation}
M_{\rm{gal}} = (5.0\pm0.1)\rm{x}10^{10} 
\left(\frac{V_c}{220~\rm{km/s}}\right)^{(3.00\pm0.07)} ~\Msun ~~.
\end{equation}
Thirdly, the external baryonic mass-fraction, $\kappa \equiv
(M_{\rm{bar,external}}/M_{\rm{gal}})$, is parameterized as a 
function of $V_c$ by a linear fit to the data shown in Fig.1, resulting in 
\begin{equation}
\kappa(V_c) = 
(1.00\pm0.07) \left(\frac{V_c}{220~\rm{km/s}}\right) - (0.31\pm0.02) ~~.
\end{equation}
The fit is valid in the range $V_c \sim$100-330 km/s, and extrapolates to
zero at $V_c$=68 km/s; in the following $\kappa$=0 is assumed for $V_c<68$
km/s, to obtain a lower limit to the global baryonic external 
mass-fraction (actually, at $V_c\la$100 km/s, $\kappa$ is expected to
start increasing with decreasing $V_c$, as SNII feedback
increasingly suppresses gas cool-out and star-formation: for a 
$\sim$3x10$^5$ particle simulation of a 
$V_c$=41 km/s galaxy, an external baryon fraction of $\sim$1.4 was found). 

The globally averaged external baryonic mass-fraction can now be expressed
as
\begin{equation}
\bar{\kappa} = \frac{
\int_0^{V_{c,\rm{max}}} \kappa(V_c)~M_{\rm{gal}}(V_c)~\Psi(V_c)~dV_c}
{\int_0^{V_{c,\rm{max}}} M_{\rm{gal}}(V_c)~\Psi(V_c)~dV_c} ~~, 
\end{equation}
where $V_{c,\rm{max}}$ is the maximum circular speed of disk galaxies.
Using eqs.[1]-[3], and assuming $V_{c,\rm{max}}$=350 km/s (disk galaxies
of such circular speeds are certainly observed), 
$\bar{\kappa}$=0.64 is obtained. Increasing $V_{c,\rm{max}}$ to 500 km/s
(which must be considered an observational upper limit), results in an increase
of $\bar{\kappa}$ to 0.72. Fitting to the various external components 
individually, shows that hot gas contributes 70\% to $\bar{\kappa}$,
inner halo stars 5\%, and outer halo stars, satellite stars and
cold gas in satellites 25\%. Assuming that the above results on external
mass fractions also apply to field elliptical/lenticular galaxies (see 
Fig.1), and applying the field galaxy velocity function of \cite{D.04},
increases the estimate of $\bar{\kappa}$ to 0.72 and 0.85, for 
$V_{c,\rm{max}}$=350 and 500 km/s, respectively (in principle, very
strong AGN feedback, not included in the current simulations, could lower
hot gas fractions of E/S0 galaxies; on the other hand E/S0s are (contrary 
to disk galaxies), in general, observed to be embedded
in hot gas haloes). Finally, since, in particular, the larger galaxies
host a number of smaller galaxies within $r_{\rm{vir}}$, the velocity
functions should actually be reduced accordingly at the low $V_c$ end. Hence,
the above $\bar{\kappa}$ estimates are, in fact, lower limits.
 
The above results, taken at face value, provide the explanation why
only about half of the baryonic mass, inside of the virial radius of
galaxies like the Milky Way, is actually found in stars and cold gas
in the galaxies. Given the ramifications of this result, it is 
important to test its robustness:
Although $f_b\sim$0.15 is indicated by
standard big bang nucleosynthesis, various CMB results and direct 
measurements of $f_b$ in galaxy clusters, lower values are still not
excluded. To test the effect of adopting $f_b=$0.10, five galaxies 
were re-simulated with this value, at the resolution of SLGP03. 
The results of these 5 simulations were compared to a sample of 39 
simulations run with $f_b=$0.15 at the resolution of SLGP03. It is
found that hot gas fractions, as well as total external baryonic
mass fractions, {\it increase} by 30-50\% (at a given $V_c$) going from
$f_b=$0.15 to 0.10. This is mainly due to the decreased efficiency of
hot gas cool-out, and the resulting increase of the virial radius at a
{\it given} circular speed of the central disk galaxy.

To test the effect of adopting $h$=0.7 and $\sigma_0$=0.9, rather than
the 0.65 and 1.0 used in the main simulations, one simulation was run
with the former parameters at the resolution of SLGP03. Comparing to
the above 39 simulations it was found that the hot gas, as well as, total
external fraction increase by $\sim$5-10\%. This is likely mainly due to
the Hubble time being 7\% less, allowing less time for hot gas cool-out.

\begin{figure}[t]
\begin{center}
\vspace*{-0.5cm}\resizebox{7.8cm}{!}{\includegraphics{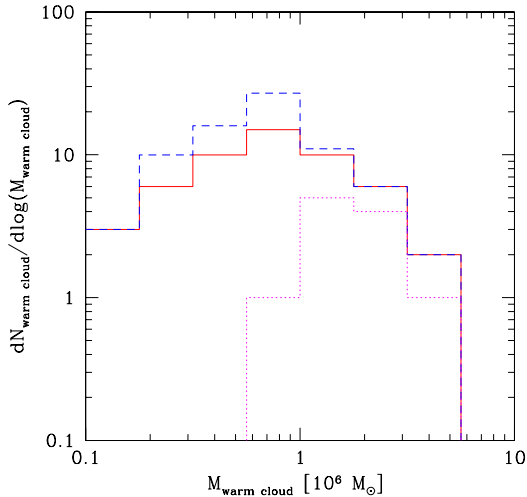}}\vspace*{-0.4cm}
\end{center}%
\vspace{-0.7cm}
\caption{Mass distribution of warm clouds at $r<$50 kpc for the ultra-high
resolution simulation at $t$=0.3 Gyr ({\it dashed line}), and 0.7 Gyr
({\it solid line}), as well as for the very high resolution simulation at 
$t$=0.7 Gyr ({\it dotted line}).}
\end{figure}
\subsection{Thermal instability of hot halo gas, and HVCs}
The hot, dilute halo gas may be susceptible to thermal instability, causing
the formation of a two-phase medium, consisting of a hot phase 
($T\sim10^6$~K) and a warm phase ($T\sim10^4$~K), in approximate pressure
equilibrium (e.g., Burkert \& Lin 2000; Maller \& Bullock 2004, Kaufmann et
al. 2005). This may not change the general picture, since warm
clouds, once formed, may quickly be destroyed again due to various physical 
processes. It is, however, important to estimate the fraction of halo gas
in warm clouds, at any given time. To estimate this for a realistic,
cosmological simulation very high numerical resolution is required, and
the following approach was adopted: for a ``standard'' high-resolution
simulation of a Milky Way like galaxy ($V_c$=224 km/s, 
$M_{\rm{vir}}$=8x10$^{11} \Msun$, $r_{\rm{vir}}$=256 kpc), at $z$=0.4, each
gas particle, at $r<$300 kpc, was split into 12 equal mass gas particles,
to achieve very high resolution of the gas phase inside of $r_{\rm{vir}}$.
The simulation was then continued for 0.2 Gyr, whence each gas particle,
at $r<$50 kpc, was again split into 8 equal mass gas particles, to
achieve ultra-high gas phase resolution in the region where most of the 
warm clouds reside (see below). 
The resulting ultra-high resolution gas particle mass is 
7.6x10$^3$~$h^{-1}\Msun$, and gravity
softening length $\epsilon_{\rm{gas}}$=83$h^{-1}$pc. The simulation was
then continued for an additional 0.5 Gyr. In Fig.4 is shown, for $r<$~50 kpc, 
the distribution of warm cloud masses, at $t$=0.3 and 0.7 Gyr. The integrated
warm cloud mass {\it decreases} somewhat with time ($\sim$20\%), to 
$\sim$5x10$^7 \Msun$ at $t$=0.7 Gyr, or about 2\% of the total mass of hot 
halo gas at $r<$~50 kpc.  Also shown, is the mass distribution of warm clouds,
at $r<$~50 kpc, in the very high resolution simulation,
at $t$=0.7 Gyr. To the cloud resolution limit of this simulation, 
$M_{\rm{cl,res}}\sim$2x10$^6 \Msun$, the agreement between the two
simulations is quite good. Moreover, with a resolution limit of  
$M_{\rm{cl,res}}\sim$3x10$^5 \Msun$ for the ultra-high resolution simulation,
the peak at $M_{\rm{cl}}\sim$7-8x10$^5 \Msun$ appears well resolved. It 
hence seems unlikely that the mass contribution from warm clouds of mass
less than about 3x10$^5 \Msun$ is significant; such clouds will in any
case quickly be destroyed by various physical processes, e.g., \cite{MB04}.

From the very high resolution simulation, the mass in warm clouds
at 50$<r<$256 kpc, is about 75\% of that within 50 kpc. Assuming this ratio,
the total mass of warm clouds, at $r<r_{\rm{vir}}$, down to 
$M_{\rm{cl}}\sim$3x10$^5 \Msun$, estimates to $\sim$10$^8 \Msun$. It seems
unlikely that inclusion of warm clouds of $M_{\rm{cl}}\la$3x$10^5 \Msun$
(which can not be resolved by the current simulation) will increase this
estimate by more than a factor of a few. \cite{P06} estimates the total
gas mass of the HVC system of the Milky Way to $\sim$4-6x10$^8~\Msun$, assuming
that the HVCs are distributed to distances of $\sim$60 kpc (a reasonable
assumption according to the present findings). This is in 
fair agreement with, though somewhat larger than, the above estimates. 
Moreover, the hot gas in the halo of the simulated Milky Way like galaxy, 
has $n_{\rm{H}}$$\sim$10$^{-3.5}$,10$^{-4}$ and 10$^{-4.5}$
cm$^{-3}$ at $r\sim$10, 50 and 100 kpc. This is in reasonable agreement with
the lower limit of $\sim$10$^{-4}$ cm$^{-3}$, deduced by \cite{QM01} from
the observed head--tail structure of many Galactic HVCs (also displayed
by many of the simulated warm clouds)$^1$.

The warm clouds are mainly confined by the pressure of the ambient halo gas,
though some of the most massive display high density cores, and have a
ratio of gravitational to thermal energy of about 0.5. Moreover,
they typically appear to be
seeded by warm filamentary structures, which are leftovers from earlier
``cold'' accretion (e.g., Birnboim \& Dekel 2003; Keres et al. 2005;
Sommer-Larsen 2005 --- details will be given in Putman \& Sommer-Larsen 2006). 
No
additional warm clouds appear to form, despite that the necessary condition
for onset of thermal instability, viz. $\tau_{\lambda}<\tau_{\rm{cool}}$, is
satisfied everywhere in the hot halo gas ($\tau_{\lambda}$ is the sound
crossing time, which is taken to be $\sim 2h_{\rm{SPH}}/c_s$, where 
$h_{\rm{SPH}}$ is the local SPH smoothing length and $c_s$ is the sound
speed: $\tau_{\rm{cool}}=\frac{E}{\dot{E}}$ is the timescale for
radiative cooling).
The mass
of hot gas within $r_{\rm{vir}}$ is $\sim$2.3x10$^{10} \Msun$, so the
ratio of warm cloud mass to hot gas mass is at most a few percent.
Although fully cosmological simulations at even higher resolution are
required to address these issues completely, the above results strongly
suggest that the hot gas phase is the dominant one in the halo. It is
also worth noting that \cite{P.06} found that the predicted halo X-ray
luminosity and surface brightness profile of large disk galaxies, simulated 
at a resolution which was, though high, not sufficiently high to resolve 
warm halo clouds, matches the observed X-ray properties of NGC5746 very
well. This would not be expected if a substantial fraction of the halo gas
was in the warm phase.

\vspace{.2cm}
I have benefited from discussions with J.~Binney, A.~Dekel, R.~Bower, 
J.~Fynbo, L.~Hernquist, T.~Kaufmann, A.~Loeb, L.~Portinari, M.~Putman 
and J.~Silk. 

The Dark Cosmology Centre is funded by the DNRF.







\clearpage


\clearpage

\end{document}